\documentclass[11pt]{article}
\usepackage{latexsym}
\usepackage{amssymb}
\usepackage[dvips]{graphicx}
\usepackage{epsfig}
\usepackage{rotating}
\usepackage{amsfonts}
\usepackage{amsmath}
\begin{document}

\begin{titlepage}
\begin{flushright}
CP3-12-34\\
ICMPA-MPA/026/2012\\
\end{flushright}

\vspace{40pt}

\begin{center}

{\Large\bf Supersymmetric Quantum Mechanics,}\\

\vspace{5pt}

{\Large\bf Engineered Hierarchies of Integrable Potentials,}\\

\vspace{5pt}

{\Large\bf and the Generalised Laguerre Polynomials}\\

\vspace{20pt}

Daddy Balondo Iyela$^{a,b,c}$,
Jan Govaerts$^{b,a,}$\footnote{Fellow of the Stellenbosch Institute
for Advanced Study (STIAS), 7600 Stellenbosch, South Africa}$^{,}$\footnote{Fellow of the Institute of Physics, UK}
and M. Norbert Hounkonnou$^{a}$

\vspace{15pt}

$^{a}${\sl International Chair in Mathematical Physics and Applications (ICMPA--UNESCO Chair),\\
University of Abomey--Calavi, 072 B. P. 50, Cotonou, Republic of Benin}\\
E-mail: {\em balondo36@gmail.com, hounkonnou@yahoo.fr, norbert.hounkonnou@cipma.uac.bj}\\
\vspace{10pt}
$^{b}${\sl Centre for Cosmology, Particle Physics and Phenomenology (CP3),\\
Institut de Recherche en Math\'ematique et Physique (IRMP),\\
Universit\'e catholique de Louvain (U.C.L.),\\
2, Chemin du Cyclotron, B-1348 Louvain-la-Neuve, Belgium}\\
E-mail: {\em Jan.Govaerts@uclouvain.be}\\
\vspace{10pt}
$^{c}${\sl D\'epartement de Physique, Universit\'e de Kinshasa (UNIKIN),\\
Kinshasa, Democratic Republic of Congo}\\

\vspace{20pt}


\vspace{10pt}

\begin{abstract}
\noindent
Within the context of Supersymmetric Quantum Mechanics and its related hierarchies of integrable quantum Hamiltonians
and potentials, a general programme is outlined and applied to its first two simplest illustrations. Going beyond the usual
restriction of shape invariance for intertwined potentials, it is suggested to require a similar relation for
Hamiltonians in the hierarchy separated by an arbitrary number of levels, $N$. By requiring further that these
two Hamiltonians be in fact identical up to an overall shift in energy, a periodic structure is installed in the
hierarchy of quantum systems which should allow for its solution. Specific classes of orthogonal polynomials
characteristic of such periodic hierarchies are thereby generated, while the methods of Supersymmetric Quantum Mechanics then lead to
generalised Rodrigues formulae and recursion relations for such polynomials. The approach also offers the practical
prospect of quantum modelling through the engineering of quantum potentials from experimental energy spectra.
In this paper these ideas are presented and solved explicitly for the cases $N=1$ and $N=2$. The latter case
is related to the generalised Laguerre polynomials, for which indeed new results are thereby obtained. At
the same time new classes of integrable quantum potentials which generalise that of the harmonic oscillator
and which are characterised by two arbitrary energy gaps are identified, for which a complete solution is achieved algebraically.
\end{abstract}

\end{center}

\end{titlepage}

\setcounter{footnote}{0}

\section{Introduction}
\label{Intro}

Factorisation methods to solve the Schr\"odinger equation are almost as old as quantum mechanics itself. Nowadays this
approach is an integral part to the techniques of Supersymmetric Quantum Mechanics (for reviews and references to the earlier literature,
see for instance Refs.\cite{SUSYQM1,SUSYQM2}). Even if only for systems with a single degree of freedom,
this is a field which still offers most tantalising and fascinating perspectives indeed, as witnessed by a renewed strong research activity
over the last decade at least\cite{Review}, which goes on unabated and keeps exploring quite many open avenues, while extensions to more degrees
of freedom, including fermionic or spin ones, is a realm remaining largely unexplored until now (for a review and further references,
see Ref.\cite{Ioffe}).

For the single degree of freedom case which is the topic of the present work, the situation may be characterized as follows
(assuming that the potential energy is such that the energy eigenspectrum be bounded below but unbounded above while also being discrete
and countable infinite without any degeneracy). Through the factorisation of the second order differential stationary Schr\"odinger equation
into the composition of two first order differential operators which are adjoints of one another, a given Hamiltonian
whose energy eigenspectrum of states would be known, is seen to belong to an infinite hierarchy of successive intertwined pairs of Hamiltonians
of which the energy eigenspectra may readily be identified starting from that of the first Hamiltonian. All of these Hamiltonians share
an identical infinite spectrum of energy eigenvalues except for the lowest lying state which is removed as one moves from one member
in the hierarchy to the next. In other words, given a single degree of freedom quantum system of which the energy eigenspectrum
is known explicitly---namely both the values of its (discrete) energy eigenvalues and the corresponding quantum states, the latter say in terms of
their (configuration space) wave functions---, there arises a semi-infinite hierarchy of integrable single degree of freedom
quantum systems all of whose energy eigenspectra are known likewise. Supersymmetric Quantum Mechanics and factorisation of the Schr\"odinger
operator thus provide an important insight leading towards the construction of nontrivial integrable quantum potentials, starting
from already known ones. Given that energy eigenstate wave functions possess a number of nodes equal to the order of the energy level
while also defining an orthonormalised countable basis of Hilbert space, the identification and classification of such integrable hierarchies
of quantum systems should also provide generally for new insights into properties of orthogonal polynomials.

However with the exception of a few cases, integrable quantum Hamiltonians are difficult to identify, to be used then as the starting member
of such integrable hierarchies. Further restrictions are required to enable the actual construction of such hierarchies.
For instance if two successive intertwined members of the hierarchy are somehow related to one another, the
induced recursion relations should allow for a solution which then identifies an integrable hierarchy of Hamiltonians. This is the
basic idea of ``shape invariance"\cite{SUSYQM1,SUSYQM2}. If the potential energies associated
to two successive Hamiltonians in the hierarchy are related simply through a redefinition of their defining parameters,
such a property of ``shape invariance of the potential energy" allows for an explicit resolution of the energy eigenspectra
of two such intertwined Hamiltonians or potential energies, hence the identification of an integrable hierarchy of quantum Hamiltonians.
Specific classifications of shape invariant potentials have been achieved in the literature\cite{SUSYQM1,SUSYQM2}, while
the known list---which includes further cases discovered since when the latter two publications have appeared\cite{Review}---is certainly
far from exhausting a complete classification (for recent developments and further references to the relevant literature, see for instance
Refs.\cite{FellowsSmith, OdakeSasaki, Ramos, PostVinet, Karadzhov}).

The motivation and results of the present work follow a similar logic, but rather than compare two successive intertwined Hamiltonians
in the hierarchy, it considers what may occur when a Hamiltonian, or potential energy, further up in the hierarchy than simply
the next one, is related to the first one in the hierarchy. More specifically, we shall be considering the restrictions
arising for the hierarchy and the potential energies defined by it, when the ground state wave functions of two members
of the hierarchy happen to coincide. This is a condition stronger than requiring shape invariance for the corresponding
two potential energies. Indeed, as is well known, from the ground state wave function one readily identifies
the potential energy, and thus implicitly also the entire energy spectrum. Hence in actual fact, if two Hamiltonians in the hierarchy
share a common quantum ground state, up to a constant shift upwards in their energy spectra they share an identical potential
energy and thus energy eigenspectrum inclusive of the wave functions for all their energy eigenstates. As a consequence,
and in a manner similar to what happens when imposing shape invariance of the potential energies,
it becomes possible to construct an integrable hierarchy of quantum systems, leading further to specific properties
for the associated orthogonal polynomials. Clearly the above restriction may somewhat be relaxed by requiring only shape invariance
rather than an exact identity for two potential energies defining two Hamiltonians separated by $N\ge 1$ steps or levels in the hierarchy,
the case $N=1$ being the only one heretofore considered in the literature in the context of shape invariant potentials.

Leaving aside for future work the general case, the present paper details the consequences of having a common
quantum ground state for a given Hamiltonian and the next-to-next one in the hierarchy, namely the case $N=2$. This is done by specifying
the energy gaps between the first two energy levels of the starting Hamiltonian, with as a result the energy eigenvalues
simply being the repeated duplication of these two gaps thus producing a two gap periodic energy spectrum. Clearly the harmonic oscillator
is one particular degenerate case of this general situation when the two energy gaps are identical, with as consequence that
two intertwined Hamiltonians in the hierarchy then share the same quantum ground state while the spectrum simply becomes
equally spaced. The latter simple case will serve the purpose of a warm-up illustration for the more general analysis which
follows when requiring identical quantum ground states for two Hamiltonians separated by $N$ levels in the
hierarchy. The cases $N=1$ and $N=2$ will explicitly be solved in this paper. The generic situation for
$N\ge 3$ is left to future work which is still on-going.

Obviously, the general case for arbitrary $N$ opens the way towards
quantum potential engineering to produce quantum systems with $N$ gap periodic energy spectra of which the $N$ first energy gaps
take specific values, predetermined for instance from experiment. The approximation of actual physical systems provided by
such engineered quantum models is expected to grow better and better as the value of $N$ keeps increasing, in a manner dependent of the values of
the $N$ first physical gaps to be reproduced. Furthermore, it is worth noting as well that such quantum potential engineering must also be
related\cite{Review} to the well established field of inverse scattering methods in integrable Hamiltonian systems and hierarchies which
allow in principle for a reconstruction of a quantum potential from the knowledge of its energy eigenvalues and phase shifts (see for instance
Ref.\cite{Das}). Finally, given the periodic structure of such hierarchies involving a finite number of characteristic classes
of wave functions and associated orthogonal polynomials, specific recursion relations between classes of
orthogonal polynomials ought to be generated from such constructions. In these two latter contexts as well, one could thus also
expect new developments following from Supersymmetric Quantum Mechanics.

The paper is organized as follows. Section 2 briefly recalls the main features of Supersymmetric Quantum Mechanics for a single degree
of freedom system on the real line, and how hierarchies of integrable quantum Hamiltonians may be constructed out of a Hamiltonian
whose energy eigenspectrum is known. Based on this understanding the programme described above in terms of periodic hierarchies is outlined.
Section 3 then applies these ideas to the simplest situation, with $N=1$, to show that the ordinary harmonic oscillator is recovered
while known properties of the associated Hermite polynomials are reproduced using the methods provided by Supersymmetric Quantum
Mechanics. Then in Section 4 we turn to the actual original content of the present work, by solving completely the $N=2$ periodic
hierarchy. Given the two gaps characteristic of that case, the corresponding potential energies are determined, as well as the
energy eigenvalues and eigenwave functions for all states in the hierarchy. It is established that the associated ensemble of orthogonal
polynomials are the generalised Laguerre polynomials. Two new generalised Rodrigues formulae are thereby discovered for these
polynomials, while the methods of Supersymmetric Quantum Mechanics provide for specific recursion relations for these functions.
Some interesting properties related to the singularities of the corresponding potentials are also commented on.
Finally in Section 5 we present our main conclusions.

\section{Engineering Hierarchies of Integrable Quantum Hamiltonians}
\label{Sect2}

\subsection{Some Basics of Supersymmetric Quantum Mechanics}
\label{Sect2.1}

Also for the purpose of establishing our notations and conventions, let us briefly outline the salient features of a given hierarchy
of intertwined quantum Hamiltonians, without providing all the justifications---readily available from
the literature\cite{SUSYQM1,SUSYQM2}---for the results stated hereafter. Levels in such a hierarchy are labelled by an index $\lambda=1,2,\cdots$
corresponding to quantum Hamiltonians $H_\lambda$, $H_1$ for $\lambda=1$ being the first member of the hierarchy.
In the configuration space representation of quantum states in terms of real wave functions $\psi(x)$ with, in the present work,
the configuration space variable $x$ assumed to take values in the real line, $x\in\mathbb{R}$, the quantum Hamiltonian operators
are expressed as,
\begin{equation}
H_\lambda=-\frac{\hbar^2}{2m_0}\,\frac{d^2}{dx^2}\,+\,V_\lambda(x),\qquad \lambda=1,2,\cdots,
\end{equation}
$V_\lambda(x)$ being the corresponding potential energies for the single degree of freedom $x$ of mass $m_0$.
The stationary Sch\"odinger equation determines the energy eigenspectrum of each system in the form,
\begin{equation}
H_\lambda \psi_{\lambda,n}(x)=E_{\lambda,n}\,\psi_{\lambda,n}(x),\qquad n=0,1,2,\cdots.
\end{equation}
The eigenwave functions are assumed to be orthonormalised,
\begin{equation}
\int_{-\infty}^{+\infty}dx\,\psi^*_{\lambda,n}(x)\,\psi_{\lambda,m}(x)=\delta_{n,m},\qquad n,m=0,1,2,\cdots,
\end{equation}
while for later purposes it is useful to consider the successive gaps in each of these energy eigenspectra,
\begin{equation}
\Delta_{\lambda,n}=E_{\lambda,n}\,-\,E_{\lambda,n-1}>0,\qquad n=1,2,3,\cdots.
\end{equation}
For convenience, let us also introduce the following change of configuration space variable, $u\in\mathbb{R}$, such that,
\begin{equation}
x=u\sqrt{\frac{\hbar^2}{2m_0}},\qquad
u=x\sqrt{\frac{2m_0}{\hbar^2}},
\end{equation}
the variable $u$ thus having the physical dimension of the inverse square root of energy, $[u]=E^{-1/2}$. The above relations
then take the form,
\begin{equation}
\left(-\frac{d^2}{du^2}\,+\,V_\lambda(u)\right)\,\psi_{\lambda,n}(u)=E_{\lambda,n}\,\psi_{\lambda,n}(u),\qquad
\int_{-\infty}^{+\infty}du\,\psi^*_{\lambda,n}(u)\,\psi_{\lambda,m}(u)=\sqrt{\frac{2m_0}{\hbar^2}}\,\delta_{n,m}
\label{eq:normal1}
\end{equation}
(with a slight customary abuse of notation for these mathematical functions).

Factorisation of the stationary Schr\"odinger equation is achieved as follows. The ground state wave function of each Hamiltonian
uniquely determines a superpotential $W_\lambda(u)$ through,
\begin{equation}
W_\lambda(u)=-\frac{1}{\psi_{\lambda,0}(u)}\,\frac{d\psi_{\lambda,0}(u)}{du}=-\frac{d}{du}\ln|\psi_{\lambda,0}(u)|.
\end{equation}
Note that this relation integrates to,
\begin{equation}
\psi_{\lambda,0}(u)=N_{\lambda,0}\,e^{-\int du\,W_\lambda(u)},
\end{equation}
$N_{\lambda,0}$ being a normalisation factor (which may be chosen to be real and positive).
In terms of $W_\lambda(u)$, one then constructs two linear first order differential operators which are adjoint of one another and
factorise the Schr\"odinger equation,
\begin{equation}
A^\dagger_\lambda=-\frac{d}{du}\,+\,W_\lambda(u),\qquad
A_\lambda=\frac{d}{du}\,+\,W_\lambda(u),
\end{equation}
and such that,
\begin{equation}
H_\lambda=A^\dagger_\lambda\,A_\lambda\,+\,E_{\lambda,0},\qquad
V_\lambda(u)=W^2_\lambda(u)\,-\,W'_\lambda(u)\,+\,E_{\lambda,0},
\end{equation}
where $W'_\lambda(u)=dW_\lambda(u)/du$ (hereafter the notation $d_u\equiv d/du$ is often used as well).

Two intertwined Hamiltonians $H_\lambda$ and $H_{\lambda + 1}$ belonging to a same hierarchy are then such that,
\begin{equation}
H_{\lambda +1}=A^\dagger_{\lambda + 1}\,A_{\lambda +1}\,+\,E_{\lambda +1,0}=A_\lambda\,A^\dagger_\lambda\,+\,E_{\lambda,0},
\end{equation}
with as consequence the following intertwining relation for the corresponding superpotentials,
\begin{equation}
V_{\lambda+1}(u)=W^2_{\lambda +1}\,-\,W'_{\lambda+1}\,+\,E_{\lambda+1,0}=W^2_\lambda\,+\,W'_\lambda\,+\,E_{\lambda,0}.
\end{equation}
Generally there is no known method for solving any of these two types of Riccati equations,
\begin{equation}
W^2_\lambda\,-\,W'_\lambda\,+\,E_{\lambda,0}=V_\lambda,\qquad
W^2_{\lambda +1}\,-\,W'_{\lambda+1}\,+\,E_{\lambda+1,0}=W^2_\lambda\,+\,W'_\lambda\,+\,E_{\lambda,0},
\end{equation}
for the superpotentials $W_\lambda(u)$ and $W_{\lambda+1}(u)$ given, say, the functions $V_\lambda(u)$ and $W_\lambda(u)$,
respectively. However knowledge of the ground state wave functions $\psi_{\lambda,0}(u)$ and $\psi_{\lambda+1,0}(u)$
provides just such a solution. As a matter of fact, knowledge of the whole energy eigenspectrum of $H_\lambda$
provides at once not only for the superpotentials $W_\lambda(u)$ but also for the whole energy eigenspectrum of $H_{\lambda+1}$,
thus in particular also $W_{\lambda+1}(u)$ from the ground state $\psi_{\lambda+1,0}(u)$. Indeed, the following relations apply between
the energy eigenspectra of the two intertwined systems which are mapped into one another by the operators $A_\lambda$ and
$A^\dagger_\lambda$ except for the ground state of $H_\lambda$ which is annnihilated by $A_\lambda$ (hereafter, $n=0,1,2,\cdots$),
\begin{equation}
\psi_{\lambda+1,n}(u)=\frac{1}{\sqrt{E_{\lambda,n+1} - E_{\lambda,0}}}\,A_\lambda\,\psi_{\lambda,n+1}(u),\quad
\psi_{\lambda,n+1}(u)=\frac{1}{\sqrt{E_{\lambda,n+1} - E_{\lambda,0}}}\,A^\dagger_\lambda\,\psi_{\lambda+1,n}(u),
\label{eq:states0}
\end{equation}
\begin{equation}
E_{\lambda+1,n}=E_{\lambda,n+1},
\end{equation}
while, by construction,
\begin{equation}
A_\lambda\,\psi_{\lambda,0}(u)=0.
\end{equation}

Consequently any such semi-infinite hierarchy of pairwise intertwined Hamiltonians is characterised by two complementary sets of information.
On the one hand, an ensemble of energy gaps\footnote{Note that since $\Delta_{\lambda+1,n}=\Delta_{\lambda,n+1}$ ($n=1,2,\cdots$),
it suffices to known the energy gaps $\Delta_{1,n}$ of the first Hamiltonian of the hierarchy, $H_1$. All energy spectra are known
from this spectrum of gaps up to the arbitrary overall shift by ground state energy of the first hierarchy, $E_{1,0}$.}
$\Delta_{\lambda=1,n}=\hbar\omega_{1,n}$ ($n=1,2,\cdots$), and on the other hand, an ensemble of intertwined superpotentials, $W_\lambda(u)$
($\lambda=1,2,\cdots$), obeying the recursion relations\footnote{Note that from this point of view, provided
these Riccati equations may be solved, it even becomes possible in principle to extend the hierarchy to negative values
of $\lambda\le 0$, hence extend the semi-infinite hierarchy to a truly infinite hierarchy of integrable quantum Hamiltonians.
Applying this remark to the harmonic oscillator as the choice for $H_1$ has produced some interesting new infinite classes of integrable
quantum systems\cite{FellowsSmith,BG}.},
\begin{equation}
W^2_{\lambda+1}(u)\,-\,W'_{\lambda+1}(u)\,+\,\Delta_{\lambda,1}=W^2_\lambda(u)\,+\,W'_\lambda(u),\qquad
\Delta_{\lambda,1}=\Delta_{1,\lambda},\qquad \lambda=1,2,\cdots.
\end{equation}
Any eigenstate $\psi_{\lambda,n}(u)$ ($n=1,2,\cdots$) of $H_\lambda$ may then be constructed from the ground state $\psi_{\lambda+n,0}(u)$
of $H_{\lambda+n}$ through the repeated application of $A^\dagger_{\lambda}$ operators defined in terms of the superpotentials,
\begin{eqnarray}
&& \psi_{\lambda,n}(u) = \nonumber \\
 &=& \left[\left(\Delta_{\lambda,n}+\Delta_{\lambda,n-1}\cdots+\Delta_{\lambda,1}\right)\cdot
\left(\Delta_{\lambda,n}+\Delta_{\lambda,n-1}+\cdots+\Delta_{\lambda,2}\right)\cdots\left(\Delta_{\lambda,n}+\Delta_{\lambda,n-1}\right)
\cdot\Delta_{\lambda,n}\right]^{-1/2}\times \nonumber \\
&&\times\ A^\dagger_\lambda\,A^\dagger_{\lambda+1}\cdots A^\dagger_{\lambda+n-1}\,\psi_{\lambda+n,0}(u),\qquad n=1,2,\cdots,
\label{eq:states1}
\end{eqnarray}
while all ground state functions are themselves directly constructed from the superpotentials,
\begin{equation}
\psi_{\lambda,0}(u)=N_{\lambda,0}\,e^{-\int du W_\lambda(u)},
\label{eq:gs1}
\end{equation}
with the normalisation factors $N_{\lambda,0}$ remaining to be determined.

\subsection{Engineering of Quantum Energy Spectra}
\label{Sect2.2}

It may so happen that two members of the hierarchy, say $H_\lambda$ and $H_{\lambda+N}$ with $N\ge 1$, share a common
ground state, or equivalently a same superpotential, $W_{\lambda+N}(u)=W_\lambda(u)$. Since these two systems then
share the same potential energy up to an overall shift in the ground state energy of $H_{\lambda+N}$ given by the sum of
the first $N$ gaps in the spectrum of $H_\lambda$, a periodic structure of order $N$ exists in the hierarchy which then
also readily extends to $-\infty<\lambda<+\infty$.

Consequently without loss of generality one may consider that the
Hamiltonians which share the same ground state are $H_1$ and $H_{1+N}$. Such an infinite hierarchy is then constructed
(up to the arbitrary energy level $E_{1,0}$) from the knowledge of $N$ gaps, namely those of\footnote{Note that
$\Delta_{\lambda,n}=\Delta_{1,n+\lambda-1}$.} $H_1$, $\Delta_n=\Delta_{1,n}=\hbar\omega_n$ for $n=1,2,\cdots,N$,
and $N$ superpotentials, namely $W_\lambda(u)$ with $\lambda=1,2,\dots,N$ since $W_{1+N}(u)=W_1(u)$, while all these finite number
of data must solve the $N$ recursion relations which close back onto themselves since we now also require that $W_{1+N}(u)=W_1(u)$,
\begin{equation}
W^2_{\lambda+1}(u)\,-\,W'_{\lambda+1}(u)\,+\,\Delta_{\lambda}=W^2_\lambda(u)\,+\,W'_\lambda(u),\qquad \lambda=1,2,\cdots,N.
\label{eq:recur1}
\end{equation}

Because of the isospectral properties of the hierarchy except for its ground states which are removed or added as one moves
up or down in the hierarchy using the operators $A_\lambda$ and $A^\dagger_\lambda$, an order $N$ periodicity arises.
Given any Hamiltonian $H_{\lambda_0}$, all its partners $H_\lambda$ at
the levels $\lambda_0$~(mod $N$), namely with $\lambda=\lambda_0+kN$ ($k\in\mathbb{Z}$), share the same energy spectrum
up to a shift in energy by $k(\Delta_1+\cdots+\Delta_N)$, and share the same energy eigenwave functions,
$\psi_{\lambda,n}(u)=\psi_{\lambda_0,n}(u)$ ($n=0,1,2,\cdots$). Furthermore,
except for their ground state energies, the gaps of each of these spectra consist of the periodic duplication of their first $N$ gaps
while preserving their order. Finally, as one considers each of the Hamiltonians $H_1$ to $H_N$ in turn, their first $N$ gaps
consist of the $N$ cyclic permutations of the $N$ gaps $(\Delta_1,\Delta_2,\cdots,\Delta_N)$ in that order, with $\Delta_1$ being
the first gap in the spectrum of $H_1$, $\Delta_2$ in that of $H_2$, and so on, until $H_{N+1}$ of which the first gap
is again $\Delta_1$.

Such a specific situation thus possibly opens the avenue towards engineering quantum potentials of which the $N$ first energy levels
take prespecified energy values, for instance dictated from experiment. The corresponding energy spectra are then $N$ gap periodic.
Such models may provide good approximations to actual physical quantum systems, with a quality of approximation that presumably
would improve as the value of $N$ increases. Once the first $N$ gaps are specified, there remains to solve the $N$ coupled nonlinear
first order differential Riccati recursion relations (\ref{eq:recur1}), for which even a numerical approach could be developed
for practical applications. Work on the generic case with $N\ge 3$ is still on-going. Hereafter the solutions for the cases
$N=1$ and $N=2$ are detailed. In a certain sense the equations to be solved then define a degeneracy of the general case when $N\ge 3$.

\section{The $N=1$ Case}
\label{Sect3}

The simplest case $N=1$ is characterised by a single gap and a single superpotential. Up to the value of their ground state
energy which is shifted by a multiple of that gap, all Hamiltonians of this hierarchy, of periodicity $N=1$, thus define
the same quantum system with an equally spaced energy spectrum. Quite obviously, all these Hamiltonians are those of the same
harmonic oscillator whose angular frequency, $\omega_0$, is set by the choice of unique energy gap. Using the notations
\begin{equation}
\Delta_1=\Delta=\hbar\omega_0,\qquad W_1(u)=W(u),\qquad E_{1,0}=E_0,
\end{equation}
the single Riccati ``recursion" relation to be solved is
\begin{equation}
W^2(u)\,-\,W'(u)\,+\,\Delta=W^2(u)\,+\,W'(u).
\end{equation}
Clearly there always exists a solution, given as,
\begin{equation}
W'(u)=\frac{1}{2}\Delta,\qquad W(u)=\frac{1}{2}\Delta\left(u-u_0\right),
\end{equation}
$u_0$ being an arbitrary integration constant. Correspondingly one finds,
\begin{eqnarray}
V_1(u) &=& W^2-W'+E_0=\frac{1}{4}\Delta^2(u-u_0)^2-\frac{1}{2}\Delta+E_0
=\frac{1}{2}m_0\omega^2_0(x-x_0)^2-\frac{1}{2}\hbar\omega_0+E_0, \nonumber \\
V_2(u) &=& W^2+W'+E_0=\frac{1}{4}\Delta^2(u-u_0)^2+\frac{1}{2}\Delta+E_0
=\frac{1}{2}m_0\omega^2_0(x-x_0)^2+\frac{1}{2}\hbar\omega_0+E_0,
\end{eqnarray}
which are indeed two choices of potential energies for the harmonic oscillator of angular frequency $\omega_0$ and
centered at $x=x_0$, shifted by its quantum energy gap $\Delta=\hbar\omega_0$.

Given (\ref{eq:gs1}), the ground state wave function inclusive of its normalisation is readily found to be,
\begin{equation}
\psi_{1,0}(u)=\left(\frac{m_0\omega_0}{\pi\hbar}\right)^{1/4}\,e^{-\frac{1}{4}\Delta(u-u_0)^2}=
\left(\frac{m_0\omega_0}{\pi\hbar}\right)^{1/4}\,e^{-\frac{m_0\omega_0}{2\hbar}(x-x_0)^2}=
\left(\frac{m_0\omega_0}{\pi\hbar}\right)^{1/4}\,e^{-\frac{1}{2}v^2},
\end{equation}
where the following change of (dimensionless) variable is introduced,
\begin{equation}
v=\sqrt{\frac{\Delta}{2}}\,(u-u_0)=\sqrt{\frac{m_0\omega_0}{\hbar}}(x-x_0)\in\mathbb{R}.
\end{equation}
The energy spectrum of the first Hamiltonian in the hierarchy, $H_1$, is simply,
\begin{equation}
E_{1,n}=E_0+n\Delta,\qquad n=0,1,2,\cdots,
\end{equation}
while the energy eigenwave functions are constructed as, given (\ref{eq:states1}),
\begin{equation}
\psi_{1,n}(u)=\frac{1}{\sqrt{\Delta^n\cdot n!}}\,\left(A^\dagger\right)^n\,\psi_{1,0}(u),\qquad n=0,1,2,\cdots,
\end{equation}
where one has,
\begin{equation}
A=A_1=d_u+W(u)=\sqrt{\frac{\Delta}{2}}\left(v\,+\,d_v\right),\qquad
A^\dagger=A^\dagger_1=-d_u+W(u)=\sqrt{\frac{\Delta}{2}}\left(v\,-\,d_v\right).
\end{equation}
Consequently, one readily finds,
\begin{equation}
\psi_{1,n}(u)=\left(\frac{m_0\omega_0}{\pi\hbar}\right)^{1/4}\,\frac{1}{\sqrt{2^n\,n!}}\,e^{-\frac{1}{2}v^2}\,P_n(v),\qquad n=0,1,2,\cdots,
\end{equation}
where $P_n(v)$ are polynomials in $v$ of order $n$ defined by the formula,
\begin{equation}
P_n(v)=e^{\frac{1}{2}v^2}\,\left(v\,-\,d_v\right)^n\,e^{-\frac{1}{2}v^2},\qquad n=0,1,2,\cdots,
\end{equation}
and obeying the following orthonormality properties, given the conditions in (\ref{eq:normal1}),
\begin{equation}
\int_{-\infty}^{+\infty}dv\,e^{-v^2}\,P_n(v)\,P_m(v)=\delta_{n,m}\,2^n\,n!\,\sqrt{\pi},\qquad n,m=0,1,2,\cdots.
\label{eq:orthoHermite}
\end{equation}
That these polynomials are precisely the usual Hermite polynomials should be quite obvious. For instance using the fact that
\begin{equation}
e^ {-\frac{1}{2}v^2}\,\left(v-d_v\right)=(-d_v)\,e^{-\frac{1}{2}v^2},
\end{equation}
implies that the above generating formula for the $P_n(v)$ polynomials is reduced to,
\begin{equation}
P_n(v)=e^{v^2}\,e^{-\frac{1}{2}v^2}\,\left(v-d_v\right)^n\,e^{-\frac{1}{2}v^2}=(-1)^n\,e^{v^2}\left(\frac{d}{dv}\right)^n\,e^{-v^2}=H_n(v),
\end{equation}
which is indeed Rodrigues' formula defining the Hermite polynomials, which also obey (\ref{eq:orthoHermite}) (see for instance Ref.\cite{GR}).

This identification may also be achieved by considering the original Schr\"odinger equation, say for $H_1$, in which the known potential energy,
energy spectrum and wave functions are substituted, namely,
\begin{equation}
\left(-d^2_u\,+\,\frac{1}{4}\Delta^2(u-u_0)^2-\frac{1}{2}\Delta+E_0\right)\,\psi_{1,n}(u)=\left(E_0+n\Delta\right)\,\psi_{1,n}(u),
\end{equation}
or equivalently,
\begin{equation}
\left(d^2_v\,-\,v^2\right)\,\psi_{1,n}=-(2n+1)\,\psi_{1,n}.
\end{equation}
In terms of the polynomials $P_n(v)$, one then finds,
\begin{equation}
\left(\frac{d^2}{dv^2}\,-2v\,\frac{d}{dv}\,+\,2n\right)\,P_n(u)=0.
\label{eq:diffHermite}
\end{equation}
This is indeed the second order linear differential equation of which the normalisable solution is proportional to
the Hermite polynomial $H_n(v)$ (the other linearly independent solution not being normalisable over $\mathbb{R}$)\cite{GR}.

Given the fact that the operators $A$ and $A^\dagger$ map these different quantum states into one another, specific recursion
relations may also be established for these polynomials by exploiting the supersymmetric quantum mechanics realisation
of the hierarchy. For instance the relation (\ref{eq:states0}) which in the present situation becomes,
\begin{equation}
\psi_{1,n+1}=\frac{1}{\sqrt{(n+1)\Delta}}\,A^\dagger\,\psi_{2,n},\qquad \psi_{2,n}=\psi_{1,n},\qquad n=0,1,2,\cdots,
\end{equation}
translates into the relation,
\begin{equation}
P_{n+1}(v)=e^{\frac{1}{2}v^2}\,\left(v\,-\,d_v\right)\,\left(e^{-\frac{1}{2}v^2}\,P_n(v)\right)=
2vP_n(v)\,-\,\frac{dP_n(v)}{dv}.
\end{equation}
However by taking the derivative of this identity and with the use of the differential equation (\ref{eq:diffHermite}), we also have,
\begin{eqnarray}
\frac{dP_n(v)}{dv} &=& \frac{d}{dv}\left(2vP_{n-1}(v) - \frac{dP_{n-1}(v)}{dv}\right) \nonumber \\
&=& 2P_{n-1}(v) + 2v\frac{dP_{n-1}}{dv}-\frac{d^2P_{n-1}}{dv^2} \nonumber \\
&=& 2n P_{n-1}(v),
\end{eqnarray}
so that the above recursion relation also writes as,
\begin{equation}
P_{n+1}(v)=2v P_n(v) - 2n P_{n-1}(v).
\end{equation}
Given the initial values $P_0(v)=1$ and, by definition, $P_{-1}(v)=0$, this is indeed the three steps defining recursion relation
for the Hermite polynomials\cite{GR}, displaying once again explicitly how the methods of Supersymmetric Quantum Mechanics provide insight
also into the properties of orthogonal polynomials. In the present case of course, nothing new in this respect is gained
from the present approach for as simple a system.

\section{The $N=2$ Case}
\label{Sect4}

The hierarchy corresponding to the case $N=2$ displays a periodicity of order $N=2$. It is characterised by two independent
gaps, $\Delta_1=\hbar\omega_1$ and $\Delta_2=\hbar\omega_2$, and two superpotentials, $W_1(u)$ and $W_2(u)$, with the restriction
that $W_3(u)=W_1(u)$ when considering the Riccati recursion equations,
\begin{eqnarray}
W^2_2 - W'_2 + \Delta_1 &=& W^2_1 + W'_1, \nonumber \\
W^2_1 - W'_1 + \Delta_2 &=& W^2_2 + W'_2.
\end{eqnarray}
Taking the sum and the difference of these two equations, one finds,
\begin{equation}
W'_2 + W'_1 = \frac{1}{2}(\Delta_2 + \Delta_1),\qquad
W^2_2 - W^2_1 = \frac{1}{2}(\Delta_2 + \Delta_1),
\end{equation}
of which the general solution is obviously,
\begin{eqnarray}
W_1(u) &=& \frac{1}{4}(\Delta_2+\Delta_1)(u-u_0) - \frac{1}{2}\frac{\Delta_2 - \Delta_1}{\Delta_2 + \Delta_1}\,\frac{1}{u-u_0}, \nonumber \\
W_2(u) &=& \frac{1}{4}(\Delta_2+\Delta_1)(u-u_0) + \frac{1}{2}\frac{\Delta_2 - \Delta_1}{\Delta_2 + \Delta_1}\,\frac{1}{u-u_0}.
\end{eqnarray}
Correspondingly the two potential energies of the hierarchy are obtained as,
\begin{eqnarray}
V_1(u)&=&\frac{1}{16}(\Delta_2 + \Delta_1)^2(u-u_0)^2 -\frac{1}{4}\frac{(\Delta_2 - \Delta_1)(\Delta_2+3\Delta_1)}{(\Delta_2+\Delta_1)^2}\,
\frac{1}{(u-u_0)^2} - \frac{1}{2}\Delta_2 + E_{1,0}, \nonumber \\
V_2(u)&=&\frac{1}{16}(\Delta_2 + \Delta_1)^2(u-u_0)^2 +\frac{1}{4}\frac{(\Delta_2 - \Delta_1)(3\Delta_2+\Delta_1)}{(\Delta_2+\Delta_1)^2}\,
\frac{1}{(u-u_0)^2} + \frac{1}{2}\Delta_1 + E_{1,0}.
\end{eqnarray}
Note how as is indeed expected, the two quantum systems $H_1$ and $H_2$ are transformed into one another under the permutation
$\Delta_1 \leftrightarrow \Delta_2$ and an upward shift in energy by $\Delta_1$ for the energy spectrum of $H_2$ as compared
to that of $H_1$. Another remark which provides a useful check on the results hereafter, is that the particular degenerate
choice $\Delta_2=\Delta_1$ indeed reproduces exactly the previous case $N=1$ as it should, namely the ordinary harmonic oscillator.

These two potentials are even in $(u-u_0)$. Unless we have the degenerate situation with $\Delta_2=\Delta_1$, both potentials
are singular infinite at $u=u_0$, one running to $(-\infty)$ and the other to $(+\infty)$ in a symmetric fashion around $u=u_0$,
depending on the sign of $(\Delta_2 - \Delta_1)$. Hence one of these potentials is then not bounded below, and yet this bottomless
throat-like potential well remains sufficiently narrow so that the spectrum of energy eigenstates is bounded below and
infinite discrete with normalisable wave functions in that case as well, as the construction hereafter establishes explicitly.

\subsection{Energy eigenspectra}
\label{Sect4.1}

For notational convenience it is useful to introduce the following parameter measuring the relative difference of the two gaps,
\begin{equation}
\alpha=\frac{1}{2}\,\frac{\Delta_2 - \Delta_1}{\Delta_2 + \Delta_1},\quad
-\frac{1}{2}<\alpha<\frac{1}{2},\quad
\frac{1}{2}+\alpha=\frac{\Delta_2}{\Delta_2 + \Delta_1},\quad
\frac{1}{2}-\alpha=\frac{\Delta_1}{\Delta_2 + \Delta_1}.
\end{equation}
The value $\alpha=0$ then corresponds to the $N=1$ hierarchy of the harmonic oscillator,
while the two quantum systems $H_1$ and $H_2$ are simply transformed into one another by changing the sign of $\alpha$.
From the two superpotentials
\begin{equation}
W_1(u)=\frac{1}{4}(\Delta_2 + \Delta_1)(u-u_0)\,-\,\frac{\alpha}{u-u_0},\qquad
W_2(u)=\frac{1}{4}(\Delta_2 + \Delta_1)(u-u_0)\,+\,\frac{\alpha}{u-u_0},
\end{equation}
and the construction (\ref{eq:gs1}), one readily finds for the two ground states,
\begin{equation}
\psi_{1,0}(u)=N_{1,0}\,|u-u_0|^\alpha\,e^{-\frac{1}{8}(\Delta_2+\Delta_1)(u-u_0)^2},\qquad
\psi_{2,0}(u)=N_{2,0}\,|u-u_0|^{-\alpha}\,e^{-\frac{1}{8}(\Delta_2+\Delta_1)(u-u_0)^2},
\end{equation}
with their normalisation factors given by,
\begin{equation}
N_{1,0}=N_0\,\frac{1}{\sqrt{\Gamma(\frac{1}{2}+\alpha)}},\qquad
N_{2,0}=N_0\,\frac{1}{\sqrt{\Gamma(\frac{1}{2}-\alpha)}},\qquad
N_0=\left(\frac{m_0(\Delta_2+\Delta_1)}{2\hbar^2}\right)^{1/4}.
\end{equation}

This form of the solutions for the ground state wave functions invites the following change of (dimensionless) variable,
\begin{equation}
v=\frac{1}{4}(\Delta_2+\Delta_1)(u-u_0)^2=\frac{1}{2}\,\frac{m_0(\Delta_2+\Delta_1)}{\hbar^2}\,(x-x_0)^2.
\end{equation}
However care must be exercised when using the inverse relations for this change of variable, since two separate situations---to be
distinguised by the notations $v_+$ and $v_-$ when necessary---must then be considered, depending on the sign of $(u-u_0)$, namely,
\begin{eqnarray}
u-u_0\ge 0&:&\ \ \ u-u_0=+\frac{2}{\sqrt{\Delta_2 + \Delta_1}}\,\sqrt{v_+},\qquad v_+\ge 0, \nonumber \\
u-u_0\le 0&:&\ \ \ u-u_0=-\frac{2}{\sqrt{\Delta_2+\Delta_1}}\,\sqrt{v_-},\qquad v_-\ge 0.
\end{eqnarray}
Most if not all expressions listed hereafter apply only in the domain $(u-u_0)\ge 0$ and for the variable $v_+$,
even though this fact will then not be emphasized explicitly and only the notation $v$ be used. When
extending expressions into the $(u-u_0)\le 0$ domain, one must beware of possible changes of sign
induced by the above change of variable in terms of $v_-$. Given these cautionary remarks, the ground state
wave functions are given as (these expressions apply as such whether in terms of $v_+$ or $v_-$),
\begin{equation}
\psi_{1,0}(u)=N_0\,\Gamma^{-1/2}(1/2+\alpha)\,v^{\frac{1}{2}\alpha}\,e^{-\frac{1}{2}v},\qquad
\psi_{2,0}(u)=N_0\,\Gamma^{-1/2}(1/2-\alpha)\,v^{-\frac{1}{2}\alpha}\,e^{-\frac{1}{2}v}.
\end{equation}

Through the construction in (\ref{eq:states1}), all other energy eigenwave functions may readily be constructed
through the repeated application of the $A^\dagger_1$ and $A^\dagger_2$ operators on these two ground state wave functions.
Furthermore, given their two gap periodicity the energy spectra of both Hamiltonians $H_1$ and $H_2$ are easily
identified. One finds, with $p=0,1,2,\cdots$,
\begin{eqnarray}
\label{eq:spectra12}
H_1 &:& E_{1,2p}=(\Delta_2+\Delta_1)\,p+E_{1,0}, \nonumber \\
&& E_{1,2p+1}=(\Delta_2+\Delta_1)\,p+\Delta_1+E_{1,0}=(\Delta_2+\Delta_1)(p+\frac{1}{2}-\alpha)+E_{1,0}, \\
H_2 &:& E_{2,2p}=(\Delta_1+\Delta_2)\,p+\Delta_1+E_{1,0}=(\Delta_2+\Delta_1)(p+\frac{1}{2}-\alpha)+E_{1,0}, \nonumber \\
&& E_{2,2p+1}=(\Delta_2+\Delta_1)(p+1)+E_{1,0}. \nonumber
\end{eqnarray}

Given the above change of variable, one now has,
\begin{equation}
A^\dagger_1=\sqrt{\Delta_2+\Delta_1}\left(\frac{1}{2}v^{1/2}-\frac{1}{2}\alpha v^{-1/2} - v^{1/2}d_v\right),\ \
A^\dagger_2=\sqrt{\Delta_2+\Delta_1}\left(\frac{1}{2}v^{1/2}+\frac{1}{2}\alpha v^{-1/2} - v^{1/2}d_v\right),
\end{equation}
as well as,
\begin{equation}
A_1=\sqrt{\Delta_2+\Delta_1}\left(\frac{1}{2}v^{1/2}-\frac{1}{2}\alpha v^{-1/2} + v^{1/2}d_v\right),\ \
A_2=\sqrt{\Delta_2+\Delta_1}\left(\frac{1}{2}v^{1/2}+\frac{1}{2}\alpha v^{-1/2} + v^{1/2}d_v\right).
\end{equation}
Note well that these expressions apply in the $v_+$ domain. In the $v_-$ domain, an extra minus sign multiplies each
of the expressions in the r.h.s. side of these four identities. Consequently all energy eigenwave functions obtained from
the action of an even number of $A^\dagger_1$ and $A^\dagger_2$ operators possess an identical expression whether in the $v_+$
or the $v_-$ domain, while all those wave functions obtained from the action of an odd number of these operators possess expressions
of opposite sign in these two domains. This observation plays an important role in the orthonormality properties of these quantum states.
Note also that one may expect that the repeated application of these operators on the wave functions defining the two ground states
would produce terms with powers of $v$ growing more and more negative as the number of operators increases, given the last two
contributions, in $v^{-1/2}$ and $v^{1/2}d_v$, in the above differential operators. However as will be established, cancellations are
such that this never happens, and indeed as ought to be expected, up to some overall factor function of $v$ including the exponential factor
$e^{-v/2}$ and a power of $v$, the wave functions are given by some polynomial in $v$ of finite order.

To see how this comes about, let us first compute some of the lowest lying states, beginning with the $H_1$ system.
A direct calculation finds,
\begin{eqnarray}
\psi_{1,1}(u) &=& \frac{1}{\sqrt{\Delta_1}}\,A^\dagger_1\,\psi_{2,0}(u)
=N_0\,\Gamma^{-1/2}(3/2-\alpha)\,v^{-\frac{1}{2}\alpha+\frac{1}{2}}\,e^{-\frac{1}{2}v}, \nonumber \\
\psi_{1,2}(u) &=& \frac{1}{\sqrt{\Delta_2(\Delta_2+\Delta_1)}}\,A^\dagger_1 A^\dagger_2\,\psi_{1,0}(u)=
N_0\,\Gamma^{-1/2}(3/2+\alpha)\,v^{\frac{1}{2}\alpha}\,e^{-\frac{1}{2}v}\,\left(v-\frac{1}{2}-\alpha\right), \nonumber \\
\psi_{1,3}(u) &=& \frac{1}{\sqrt{\Delta_1(\Delta_1+\Delta_2)(\Delta_1+\Delta_2+\Delta_1)}}\,A^\dagger_1 A^\dagger_2 A^\dagger_1\,\psi_{2,0}(u) \\
&=& N_0\,\Gamma^{-1/2}(5/2-\alpha)\,v^{-\frac{1}{2}\alpha+\frac{1}{2}}\,e^{-\frac{1}{2}v}\,\left(v-\frac{3}{2}+\alpha\right). \nonumber
\end{eqnarray}
Likewise for $H_2$, a direct calculation of $\psi_{2,1}(u)$, $\psi_{2,2}(u)$ and $\psi_{2,3}(u)$ using the same approach
finds the same final expressions as above with simply the substitution of $\alpha$ by $(-\alpha)$ everywhere, namely,
\begin{eqnarray}
\psi_{2,1}(u) &=& \frac{1}{\sqrt{\Delta_2}}\,A^\dagger_2\,\psi_{1,0}(u)
=N_0\,\Gamma^{-1/2}(3/2+\alpha)\,v^{\frac{1}{2}\alpha+\frac{1}{2}}\,e^{-\frac{1}{2}v}, \nonumber \\
\psi_{2,2}(u) &=& \frac{1}{\sqrt{\Delta_1(\Delta_1+\Delta_2)}}\,A^\dagger_2 A^\dagger_1\,\psi_{2,0}(u)=
N_0\,\Gamma^{-1/2}(3/2-\alpha)\,v^{-\frac{1}{2}\alpha}\,e^{-\frac{1}{2}v}\,\left(v-\frac{1}{2}+\alpha\right), \nonumber \\
\psi_{2,3}(u) &=& \frac{1}{\sqrt{\Delta_2(\Delta_2+\Delta_1)(\Delta_2+\Delta_1+\Delta_2)}}\,A^\dagger_2 A^\dagger_1 A^\dagger_2\,\psi_{1,0}(u) \\
&=& N_0\,\Gamma^{-1/2}(5/2+\alpha)\,v^{\frac{1}{2}\alpha+\frac{1}{2}}\,e^{-\frac{1}{2}v}\,\left(v-\frac{3}{2}-\alpha\right). \nonumber
\end{eqnarray}
In these calculations the cancellation mechanism mentioned above is already operational starting with the second excited level, $n=2$.
This is a general feature which extends to all excited states.

A careful analysis for the other energy eigenstates uncovers the following structure\footnote{Once again let us recall that
these expressions apply in the $v_+$ domain. In the $v_-$ domain, those for $\psi_{1,2p+1}(u)$ and $\psi_{2,2p+1}(u)$
acquire an overall additional minus sign, while those for $\psi_{1,2p}(u)$ and $\psi_{2,2p}(u)$ remain unaffected.}.
For the Hamiltonian $H_1$, one finds, for $p=0,1,2,\cdots$,
\begin{eqnarray}
\psi_{1,2p}(u) &=& \frac{N_0}{\sqrt{p!\,\Gamma(p+1/2+\alpha)}}\,v^{\frac{1}{2}\alpha}\,e^{-\frac{1}{2}v}\,P^{\rm even}_{1,p}(v), \nonumber \\
\psi_{1,2p+1}(u) &=& \frac{N_0}{\sqrt{p!\,\Gamma(p+3/2-\alpha)}}\,v^{\frac{1}{2}-\frac{1}{2}\alpha}\,e^{-\frac{1}{2}v}\,
P^{\rm odd}_{1,p}(v),
\label{eq:psi1}
\end{eqnarray}
where the polynomial factors are obtained from,
\begin{eqnarray}
P^{\rm even}_{1,p}(v) &=& v^{-\frac{1}{2}\alpha}\,e^{\frac{1}{2}v}\,D^p(\alpha)\,v^{\frac{1}{2}\alpha}\,e^{-\frac{1}{2}v}, \nonumber \\
P^{\rm odd}_{1,p}(v) &=& v^{-\frac{1}{2}+\frac{1}{2}\alpha}\,e^{\frac{1}{2}v}\,D^p(\alpha)\,
\left(\frac{1}{2}v^{1/2}-\frac{1}{2}\alpha v^{-1/2}-v^{1/2}d_v\right)\,v^{-\frac{1}{2}\alpha}\,e^{-\frac{1}{2}v} \nonumber \\
&=& v^{-\frac{1}{2}+\frac{1}{2}\alpha}\,e^{\frac{1}{2}v}\,D^p(\alpha)\, v^{\frac{1}{2}-\frac{1}{2}\alpha}\,e^{-\frac{1}{2}v},
\end{eqnarray}
in terms of the nonlinear second order differential operator $D(\alpha)$ defined by,
\begin{equation}
D(\alpha)=\left(\frac{1}{2}v^{1/2}-\frac{1}{2}\alpha v^{-1/2} - v^{1/2}d_v\right)
\left(\frac{1}{2}v^{1/2}+\frac{1}{2}\alpha v^{-1/2} - v^{1/2}d_v\right).
\end{equation}
Likewise, a direct analysis confirms again the $(\alpha\leftrightarrow -\alpha)$ substitution rule for passing from the $H_1$
system to the $H_2$ one. One finds, for $p=0,1,2,\cdots$,
\begin{eqnarray}
\psi_{2,2p}(u) &=& \frac{N_0}{\sqrt{p!\,\Gamma(p+1/2-\alpha)}}\,v^{-\frac{1}{2}\alpha}\,e^{-\frac{1}{2}v}\,P^{\rm even}_{2,p}(v), \nonumber \\
\psi_{2,2p+1}(u) &=& \frac{N_0}{\sqrt{p!\,\Gamma(p+3/2+\alpha)}}\,v^{\frac{1}{2}+\frac{1}{2}\alpha}\,e^{-\frac{1}{2}v}\,
P^{\rm odd}_{2,p}(v),
\end{eqnarray}
where,
\begin{eqnarray}
P^{\rm even}_{2,p}(v) &=& v^{\frac{1}{2}\alpha}\,e^{\frac{1}{2}v}\,D^p(-\alpha)\,v^{-\frac{1}{2}\alpha}\,e^{-\frac{1}{2}v}, \nonumber \\
P^{\rm odd}_{2,p}(v) &=& v^{-\frac{1}{2}-\frac{1}{2}\alpha}\,e^{\frac{1}{2}v}\,D^p(-\alpha)\,
\left(\frac{1}{2}v^{1/2}+\frac{1}{2}\alpha v^{-1/2}-v^{1/2}d_v\right)\,v^{\frac{1}{2}\alpha}\,e^{-\frac{1}{2}v} \nonumber \\
&=& v^{-\frac{1}{2}-\frac{1}{2}\alpha}\,e^{\frac{1}{2}v}\,D^p(-\alpha)\, v^{\frac{1}{2}+\frac{1}{2}\alpha}\,e^{-\frac{1}{2}v}.
\end{eqnarray}

The previous explicit examples thus correspond to,
\begin{eqnarray}
P^{\rm even}_{1,0}=1,\quad P^{\rm even}_{1,1}(v)=v-\frac{1}{2}-\alpha &;&
P^{\rm even}_{2,0}(v)=1,\quad P^{\rm even}_{2,1}(v)=v-\frac{1}{2}+\alpha, \nonumber \\
P^{\rm odd}_{1,0}(v)=1,\quad P^{\rm odd}_{1,1}(v)=v-\frac{3}{2}+\alpha &;&
P^{\rm odd}_{2,0}(v)=1,\quad P^{\rm odd}_{2,1}(v)=v-\frac{3}{2}-\alpha.
\end{eqnarray}
Let us compare these expressions with those for the generalised Laguerre polynomials defined by their Rodrigues formula,
\begin{equation}
L^{(\gamma)}_n(x)=\frac{1}{n!}\,x^{-\gamma}\,e^x\,\frac{d^n}{dx^n}\left(x^{n+\gamma}\,e^{-x}\right),\qquad n=0,1,2,\cdots,
\end{equation}
with in particular,
\begin{equation}
L^{(\gamma)}_0(x)=1,\qquad
L^{(\gamma)}_1(x)=(1+\gamma)-x,\qquad
L^{(\gamma)}_2(x)=\frac{1}{2}(\gamma+2)(\gamma+1) - (\gamma+2) x + \frac{1}{2}x^2.
\end{equation}
Hence we may certainly write,
\begin{eqnarray}
P^{\rm even}_{1,0}(v) &=& L^{(-\frac{1}{2}+\alpha)}_0(v),\qquad P^{\rm even}_{1,1}(v)=-L^{(-\frac{1}{2}+\alpha)}_1(v), \nonumber \\
P^{\rm odd}_{1,0}(v) &=& L^{(\frac{1}{2}-\alpha)}_0(v),\qquad P^{\rm odd}_{1,1}(v)=-L^{(\frac{1}{2}-\alpha)}_1(v),
\end{eqnarray}
and likewise for $P^{\rm even}_{2,0}(v)$, $P^{\rm even}_{2,1}(v)$, $P^{\rm odd}_{2,0}(v)$ and $P^{\rm odd}_{2,1}(v)$
under the substitution $(\alpha\rightarrow -\alpha)$. These observations
thus suggest the following general identification\footnote{The factor $p!$ and the sign $(-1)^p$ follow from considering the factor multiplying
the highest power in $v$ as generated for both classes of polynomials from their respective formulae above.} (in the $v_+$ domain),
\begin{equation}
P^{\rm even}_{1,p}(v)=(-1)^p\,p!\,L^{(-\frac{1}{2}+\alpha)}_p(v),\qquad
P^{\rm odd}_{1,p}(v)=(-1)^p\,p!\,L^{(\frac{1}{2}-\alpha)}_p(v),
\end{equation}
\begin{equation}
P^{\rm even}_{2,p}(v)=(-1)^p\,p!\,L^{(-\frac{1}{2}-\alpha)}_p(v),\qquad
P^{\rm odd}_{2,p}(v)=(-1)^p\,p!\,L^{(\frac{1}{2}+\alpha)}_p(v).
\end{equation}

That these identifications are indeed correct will be confirmed hereafter. Assuming this to be the case, in conclusion we have
thus obtained for both $H_1$ and $H_2$ their spectra, see (\ref{eq:spectra12}), but also their energy eigenwave functions, in the form,
\begin{eqnarray}
\psi_{1,2p}(u) &=& (-1)^p\,N_0\,\sqrt{\frac{p!}{\Gamma(p+1/2+\alpha)}}
\,v^{\frac{1}{2}\alpha}\,e^{-\frac{1}{2}v}\,L^{(-\frac{1}{2}+\alpha)}_p(v), \nonumber \\
\psi_{1,2p+1}(u) &=& (-1)^p\,N_0\,\sqrt{\frac{p!}{\Gamma(p+3/2-\alpha)}}
\,v^{\frac{1}{2}-\frac{1}{2}\alpha}\,e^{-\frac{1}{2}v}\,L^{(\frac{1}{2}-\alpha)}_p(v),
\label{eq:sol1}
\end{eqnarray}
\begin{eqnarray}
\psi_{2,2p}(u) &=& (-1)^p\,N_0\,\sqrt{\frac{p!}{\Gamma(p+1/2-\alpha)}}
\,v^{-\frac{1}{2}\alpha}\,e^{-\frac{1}{2}v}\,L^{(-\frac{1}{2}-\alpha)}_p(v), \nonumber \\
\psi_{2,2p+1}(u) &=& (-1)^p\,N_0\,\sqrt{\frac{p!}{\Gamma(p+3/2+\alpha)}}
\,v^{\frac{1}{2}+\frac{1}{2}\alpha}\,e^{-\frac{1}{2}v}\,L^{(\frac{1}{2}+\alpha)}_p(v),
\label{eq:sol2}
\end{eqnarray}
where $N_0=\left(m_0(\Delta_2+\Delta_1)/(2\hbar^2)\right)^{1/4}$. In particular, note how in the limit when $\Delta_2=\Delta_1$,
namely $\alpha=0$, all these results do reproduce those for the harmonic oscillator of angular frequency $\omega_0=(\Delta_2+\Delta_1)/(2\hbar)$,
given the identities\cite{GR},
\begin{equation}
H_{2p}(x)=(-1)^p\,2^{2p}\,p!\, L^{(-\frac{1}{2})}_p(x^2),\qquad
H_{2p+1}(x)=(-1)^p\,2^{2p+1}\,p!\,x\,L^{(\frac{1}{2})}_p(x^2).
\end{equation}

That the orthonormality properties (\ref{eq:normal1}) are also obtained follows from the identity\cite{GR},
\begin{equation}
\int_0^{\infty}dx\,e^{-x}\,x^{\gamma}\,L^{(\gamma)}_n(x)\,L^{(\gamma)}_m(x)=
\delta_{n,m}\,\frac{1}{n!}\,\Gamma(\gamma+n+1),\qquad n,m=0,1,2,\cdots.
\label{eq:overlap}
\end{equation}
For instance for the $H_1$ system (and likewise for the $H_2$ system), through a careful analysis of the contributions
from the $v_+$ and $v_-$ domains the normalisation of the states $\psi_{1,2p}(u)$ and $\psi_{1,2p+1}(u)$ requires, respectively,
\begin{eqnarray}
\int_0^\infty\,dv\,e^{-v}\,v^{-\frac{1}{2}+\alpha}\left(L^{(-\frac{1}{2}+\alpha)}_p(v)\right)^2 &=& \frac{1}{p!}\,\Gamma(p+1/2+\alpha), \nonumber \\
\int_0^\infty\,dv\,e^{-v}\,v^{\frac{1}{2}-\alpha}\left(L^{(\frac{1}{2}-\alpha)}_p(v)\right)^2 &=& \frac{1}{p!}\,\Gamma(p+3/2-\alpha),
\end{eqnarray}
which is indeed correct. Similarly the overlaps between two ``even" states $\psi_{1,2p}(u)$ (or $\psi_{2,2p}(u)$) on the one hand,
or two ``odd" states $\psi_{1,2p+1}(u)$ (or $\psi_{2,2p+1}(u)$) on the other hand, for two different values of $p$, are directly seen
to be vanishing as they should, given the result (\ref{eq:overlap}). Finally, the overlap between an ``even" and an ``odd" state
for a same Hamiltonian, $H_1$ or $H_2$, is found to vanish as well, on account of the specific changes of sign that arise when considering
the overlap integral over $u\in\mathbb{R}$ decomposed in terms of the two domains $v_+$ and $v_-$.

\subsection{Orthogonal polynomials generated by the $N=2$ hierarchy}
\label{Sect4.2}

In order to determine which are the orthogonal polynomials $P^{\rm even}_{1,p}(v)$ and $P^{\rm odd}_{1,p}(v)$ (and
$P^{\rm even}_{2,p}(v)$ and $P^{\rm odd}_{2,p}(v)$) generated by this $N=2$ construction of supersymmetric quantum
Hamiltonian hierarchies, let us turn to the stationary Schr\"odinger equation defined by $H_1$ (or $H_2$),
since both the energy eigenvalues and eigenwave functions are known. Given the potential energy $V_1(u)$
determined previously, and in terms of the change of variable $v$ or $v_\pm$, the stationary Schr\"odinger equation
for\footnote{The same detailed analysis for $H_2$ leads to the same conclusion, and amounts simply to the substitution everywhere of
$\alpha\rightarrow -\alpha$.} $H_1$ becomes,
\begin{equation}
\left(-vd^2_v -\frac{1}{2}d_v +\frac{1}{4}v+\frac{1}{4}\frac{\alpha(\alpha-1)}{v}-\frac{1}{2}\left(\alpha+\frac{1}{2}\right)\right)
\,\psi_{1,n}(u)=\frac{(E_{1,n}-E_{1,0})}{\Delta_2+\Delta_1}\,\psi_{1,n}(u).
\end{equation}
Distinguishing now the two situations when $n=2p$ or $n=2p+1$, on account of the expressions in (\ref{eq:psi1})
one finds for the relevant polynomials,
\begin{equation}
\left(v\frac{d^2}{dv^2}+\left(\alpha-\frac{1}{2}-v+1\right)\frac{d}{dv} + p\right)P^{\rm even}_{1,p}(v)=0,
\end{equation}
\begin{equation}
\left(v\frac{d^2}{dv^2}+\left(\frac{1}{2}-\alpha-v+1\right)\frac{d}{dv}+p\right)P^{\rm odd}_{1,p}(v)=0.
\end{equation}
Since the generalised Laguerre polynomials $L^{(\gamma)}_n(x)$ are the normalisable solutions (for the measure
$dx\,e^{-x} x^\gamma$) to the linear second order differential equation\cite{GR}
\begin{equation}
\left(x\frac{d^2}{dx^2}+(\gamma-x+1)\frac{d}{dx}+n\right)\,L(x)=0,\qquad n=0,1,2,\cdots,
\label{eq:diffL2}
\end{equation}
indeed, up to a normalisation factor that has been determined previously already, the polynomials $P^{\rm even}_{1,p}(v)$
and $P^{\rm odd}_{1,p}(v)$ coincide with the generalised Laguerre polynomials $L^{(-\frac{1}{2}+\alpha)}_p(v)$ and
$L^{(\frac{1}{2}-\alpha)}_p(v)$, respectively. The previous conjectured identification is thus established.
The $N=2$ hierarchy of integrable quantum Hamiltonians as constructed following the programme outlined in the
present work is indeed a hierarchy generating the generalised Laguerre polynomials as the associated set of
orthogonal polynomials.

\subsection{A generalised Rodrigues formula for generalised Laguerre polynomials}
\label{Sect4.3}

Given the identification of orthogonal polynomials having been achieved from the $H_1$ system, we have thus established
the following formulae for generalised Laguerre polynomials\footnote{The formulae obtained from the $H_2$ system are
the same under the substitution $(\alpha\rightarrow -\alpha)$.},
\begin{eqnarray}
L^{(-\frac{1}{2}+\alpha)}_p(v) &=& \frac{(-1)^p}{p!}\,v^{-\frac{1}{2}\alpha}\,e^{\frac{1}{2}v}\,
D^p(\alpha)\,v^{\frac{1}{2}\alpha}\,e^{-\frac{1}{2}v}, \nonumber \\
L^{(\frac{1}{2}-\alpha)}_p(v) &=& \frac{(-1)^p}{p!}\,v^{-\frac{1}{2}+\frac{1}{2}\alpha}\,e^{\frac{1}{2}v}\,
D^p(\alpha)\,v^{\frac{1}{2}-\frac{1}{2}\alpha}\,e^{-\frac{1}{2}v}.
\end{eqnarray}
By introducing a parameter $\gamma$ defined to be either $\gamma=-1/2+\alpha$ (given the first identity) or
$\gamma=1/2-\alpha$ (given the second identity), in either case these two formulae are reduced to the following two
generating formulae for generalised Laguerre polynomials,
\begin{equation}
L^{(\gamma)}_p(v) = \frac{(-1)^p}{p!}\,v^{-\frac{1}{4}-\frac{1}{2}\gamma}\,e^{\frac{1}{2}v}\,
D^p_2(\pm\gamma)\,v^{\frac{1}{4}+\frac{1}{2}\gamma}\,e^{-\frac{1}{2}v},
\label{eq:rodrigues2}
\end{equation}
where the following two nonlinear second order differential operators are introduced,
\begin{equation}
D_2(\pm\gamma)=\left(\frac{1}{2}v^{1/2}-\frac{1}{2}\left(\frac{1}{2}\pm\gamma\right)v^{-1/2}-v^{1/2}d_v\right)
\left(\frac{1}{2}v^{1/2}+\frac{1}{2}\left(\frac{1}{2}\pm\gamma\right)v^{-1/2}-v^{1/2}d_v\right).
\end{equation}
Note that these same operators may also be expressed as,
\begin{equation}
D_2(\pm\gamma)=v\left(\frac{1}{2}-\frac{1}{2}\left(\frac{3}{2}\pm\gamma\right)\frac{1}{v}-d_v\right)
\left(\frac{1}{2}+\frac{1}{2}\left(\frac{1}{2}\pm\gamma\right)\frac{1}{v}-d_v\right).
\end{equation}

As compared to the usual Rodrigues formula for generalised Laguerre polynomials,
\begin{equation}
L^{(\gamma)}_p(v)=\frac{1}{p!}\,e^v\,v^{-\gamma}\,\left(\frac{d}{dv}\right)^p\,e^{-v}\,v^{p+\gamma},
\label{eq:rodrigues}
\end{equation}
the two new generalised Rodrigues formulae for generalised Laguerre polynomials in (\ref{eq:rodrigues2})
are noteworthy. First note that irrespective of whether one uses the operator $D_2(\gamma)$ or the
operator $D_2(-\gamma)$, the same generalised Laguerre polynomial is generated from (\ref{eq:rodrigues2}).
Such a very specific feature can only apply provided once again a series of cancellations occur when expanding
the action of the powers of these operators. This may indeed be confirmed by a direct evaluation of some
of the polynomials of lowest order.

A second noteworthy property of the formulae in (\ref{eq:rodrigues2}) is the following. In contradistinction to
the usual Rodrigues formula (\ref{eq:rodrigues}) which makes it obvious that the $p$ order action of the linear first
order differential operator $d/dv$ generates a polynomial in $v$ of order $p$, the formulae in (\ref{eq:rodrigues})
may naively seem to generate some polynomial in $v$ of order $2p$ because of the $p$ order action of the nonlinear
second order differential operators $D_2(\pm\gamma)$ of which the leading term is in $v$. Yet this not the case,
once again owing to specific cancellations that occur such that no negative powers of $v$ remain either in spite of the contributions
in $1/v$ in $D_2(\pm\gamma)$, while also keeping the polynomial of order $p$ in $v$ and not higher.

In these two respects the formulae in (\ref{eq:rodrigues2}) are indeed totally different from the usual
Rodrigues formula for generalised Laguerre polynomials. To the authors' best knowledge, the generalised
Rodrigues formulae in (\ref{eq:rodrigues2}) are new in the literature. And they indeed follow directly from
the structure of the $N=2$ hierarchy of integrable quantum Hamiltonians constructed along the lines of the
programme outlined in this work.

\subsection{Recursion relations}
\label{Sect4.4}

As a first way of generating a recursion relation between polynomials of different order in $v$, consider
the formulae (\ref{eq:rodrigues2}) in the following form,
\begin{eqnarray}
L^{(\gamma)}_{p+1} &=& \frac{(-1)^{p+1}}{(p+1)!}\,v^{-\frac{1}{4}-\frac{1}{2}\gamma}\,e^{\frac{1}{2}v}\,
D_2(\pm\gamma)\,v^{\frac{1}{4}+\frac{1}{2}\gamma}\,e^{-\frac{1}{2}v}\,v^{-\frac{1}{4}-\frac{1}{2}\gamma}\,
e^{\frac{1}{2}v}\,D^p_2(\pm\gamma)\,v^{\frac{1}{4}+\frac{1}{2}\gamma}\,e^{-\frac{1}{2}v} \nonumber \\
&=& -\frac{1}{p+1}\,v^{-\frac{1}{4}-\frac{1}{2}\gamma}\,e^{\frac{1}{2}v}\,D_2(\pm\gamma)
\left(v^{\frac{1}{4}+\frac{1}{2}\gamma}\,e^{-\frac{1}{2}v}\,L^{(\gamma)}_p(v)\right).
\end{eqnarray}
An explicit evaluation of the latter expression then finds, once again following the cancellation of a number of terms,
\begin{equation}
L^{(\gamma)}_{p+1}(v)=\frac{1}{p+1}\left((\gamma+1-v)\,L^{(\gamma)}_p(v)\,-\,
(\gamma+1-2v)\,d_v L^{(\gamma)}_p(v)\,-\,vd^2_v L^{(\gamma)}_p(v)\right).
\end{equation}
However by using (\ref{eq:diffL2}) to substitute for $vd^2_vL^{(\gamma)}_p(v)$ in this identity, one has,
\begin{equation}
L^{(\gamma)}_{p+1}(v)=\frac{1}{p+1}\left((p+1+\gamma-v)\,L^{(\gamma)}_p(v)\,+\,v d_v L^{(\gamma)}_p(v)\right),
\label{eq:recur0}
\end{equation}
which is an already known recursion relation for these polynomials\cite{GR}. Nevertheless, this conclusion confirms once
again the correct identification made for the polynomials related to this $N=2$ hierarchy construction, inclusive
of their normalisation factors and choices of sign.

Since all the energy eigenwave functions of both Hamiltonians $H_1$ and $H_2$ defining the $N=2$ hierarchy are
expressed in terms of generalised Laguerre polynomials, through the action of the operators $A_1$, $A_2$, $A^\dagger_1$
and $A^\dagger_2$ on these states it becomes possible also to derive a series of recursion relations for these
polynomials. Possibly some of these relations could be new. This is to be done by considering the general relations
in (\ref{eq:states0}) and using all four possible actions of the operators mapping between levels within the hierarchy,
whether the excitation level $n$ is even or odd. A detailed and careful analysis of all these possibilities leads to
the following four independent relations, after the introduction of the relevant parameter $\gamma$ as a function of $\alpha$
depending on which case is being considered,
\begin{eqnarray}
L^{(\gamma-1)}_p(v) &=& \frac{1}{p+\gamma}\,v^{\frac{1}{4}-\frac{1}{2}\gamma}\,e^{\frac{1}{2}v}\,
\left(\frac{1}{2}v^{1/2}-\frac{1}{2}\left(\frac{1}{2}-\gamma\right) v^{-1/2} + v^{1/2} d_v\right)\,
v^{\frac{1}{4}+\frac{1}{2}\gamma}\,e^{-\frac{1}{2}v}\,L^{(\gamma)}_p(v), \nonumber \\
L^{(\gamma+1)}_p(v) &=& -\,v^{-\frac{3}{4}-\frac{1}{2}v}\,e^{\frac{1}{2}v}\,
\left(\frac{1}{2}v^{1/2}-\frac{1}{2}\left(\frac{1}{2}+\gamma\right) v^{-1/2}+v^{1/2}d_v\right)\,
v^{\frac{1}{4}+\frac{1}{2}\gamma}\,e^{-\frac{1}{2}v}\,L^{(\gamma)}_{p+1}(v), \nonumber \\
L^{(\gamma+1)}_p(v) &=& v^{-\frac{3}{4}-\frac{1}{2}\gamma}\,e^{\frac{1}{2}v}\,
\left(\frac{1}{2}v^{1/2}+\frac{1}{2}\left(\frac{1}{2}+\gamma\right) v^{-1/2} - v^{1/2}d_v\right)\,
v^{\frac{1}{4}+\frac{1}{2}\gamma}\,e^{-\frac{1}{2}v}\,L^{(\gamma)}_p(v), \\
L^{(\gamma-1)}_{p+1}(v) &=& -\frac{1}{p+1}\,v^{\frac{1}{4}-\frac{1}{2}\gamma}\,e^{\frac{1}{2}v}\,
\left(\frac{1}{2}v^{1/2}+\frac{1}{2}\left(\frac{1}{2}-\gamma\right) v^{-1/2} - v^{1/2}d_v\right)\,
v^{\frac{1}{4}+\frac{1}{2}\gamma}\,e^{-\frac{1}{2}v}\,L^{(\gamma)}_p(v). \nonumber
\end{eqnarray}
Working out these expressions explicitly, once again cancellations occur and one is left with,
\begin{eqnarray}
L^{(\gamma-1)}_p(v) &=& \frac{1}{p+\gamma}\left(\gamma + v d_v\right) L^{(\gamma)}_p(v), \nonumber \\
L^{(\gamma+1)}_p(v) &=& - d_v L^{(\gamma)}_{p+1}(v), \nonumber \\
L^{(\gamma+1)}_p(v) &=& (1-d_v) L^{(\gamma)}_p(v) , \\
L^{(\gamma-1)}_{p+1}(v) &=& -\frac{1}{p+1}\left(v-\gamma -vd_v\right) L^{(\gamma)}_p(v). \nonumber
\end{eqnarray}
Using for instance the second of these relations, the other three become,
\begin{eqnarray}
L^{(\gamma-1)}_p(v) &=& \frac{1}{p+\gamma}\left(\gamma L^{(\gamma)}_p(v) - v L^{(\gamma+1)}_{p-1}(v)\right), \nonumber \\
L^{(\gamma+1)}_p(v) &=& L^{(\gamma)}_p(v) + L^{(\gamma+1)}_{p-1}(v), \\
L^{(\gamma-1)}_{p+1}(v) &=& -\frac{1}{p+1}\left((v-\gamma) L^{(\gamma)}_p(v) + v L^{(\gamma+1)}_{p-1}(v)\right),
\end{eqnarray}
while (\ref{eq:recur0}) then reduces to another three terms recursion relation,
\begin{equation}
L^{(\gamma)}_{p+1}(v) = \frac{1}{p+1}
\left((p+1+\gamma-v) L^{(\gamma)}_p(v) -v L^{(\gamma+1)}_{p-1}(v)\right).
\end{equation}
By taking different linear combinations of these recursion relations, one reproduces recursion relations for generalised
Laguerre polynomials known from the literature and standard Tables\cite{GR}.

\subsection{Further comments}
\label{Sect4.5}

Given the above complete resolution of the $N=2$ case for a periodic hierarchy of integrable quantum Hamiltonians using
the methods of Supersymmetric Quantum Mechanics, a few more remarks are probably useful and relevant.

Among the examples of shape invariant potentials and their corresponding superpotentials listed on pages 40 and 41
of Ref.\cite{SUSYQM2}, the second one is that of a superpotential of the form (in units such that $2m_0=1$ and $\hbar=1$),
\begin{equation}
W(r)=\frac{1}{2}\omega\,r\,-\,\frac{\ell+1}{r},\qquad
V(r)=\frac{1}{4}\omega^2 r^2+\frac{\ell(\ell+1)}{r^2}-\left(\ell+\frac{3}{2}\right)\omega,\qquad E_{0,1}=0,
\label{eq:shape}
\end{equation}
$r>0$ being the radial coordinate of the three dimensional spherically symmetric harmonic oscillator,
and $\ell=0,1,2,\cdots$ the usual orbital angular-momentum quantum number, leading to an equally spaced spectrum
of eigenvalues, $E_{1,n}=2n\omega$ ($n=0,1,2,\cdots$). With the correspondences $\omega=(\Delta_2+\Delta_1)/2$ and $\ell+1=\alpha$,
it may appear that our $N=2$ construction would coincide with that situation. But this is not the case.
For one thing our general solution presents a periodic two gap structure, and is not equally spaced in its energy spectrum.
Secondly, the variable $u\in\mathbb{R}$ takes values in the entire real line, and is not a radial variable as is $r>0$.
In particular, the integration measure over the configuration space coordinate which enters the orthonormalisation
conditions for wave functions is thereby bound to be different as well, since indeed that example of a shape invariant
potential applies to the three dimensional harmonic oscillator. And finally, the ranges of values for $(\ell+1\ge 1)$
in one case, and $(-1/2<\alpha<+1/2)$ in our case, are different.

A further difference with that example from Ref.\cite{SUSYQM2} is that while the potential following from (\ref{eq:shape})
is bounded below even though divergent at the finite value $r=0$ (with the range $0<r<\infty$), in our case among the two
potential energies $V_1$ and $V_2$ one of these always is unbounded below---which of the two being
dependent\footnote{$V_1$ is unbounded below for $\alpha>0$, while it is $V_2$ when $\alpha<0$.} on the sign of $\alpha$
or $(\Delta_2 - \Delta_1)$---while in both cases the potential energy diverges at $u=u_0$. And even though
all energy eigenwave functions remain normalisable for both potentials, these features imply specific behaviour of the
wave functions around the singularity at $u=u_0$, which in our case may be approached from both sides on the real line.

A careful consideration of the complete solutions in (\ref{eq:sol1}) and (\ref{eq:sol2}) finds that nothing par\-ti\-cu\-lar
occurs for the ``odd" states $\psi_{1,2p+1}(u)$ or $\psi_{2,2p+1}(u)$. Given the range $(-1/2<\alpha<+1/2)$,
these wave functions remain finite at $u=u_0$, and in fact even vanish at that point (which, of course, is consistent
with the sign issue for the two domains $v_+$ and $v_-$, and the fact that the potentials are even in $(u-u_0)$ while
the ``odd" states are odd functions of that variable).
Given that the generalised Laguerre polynomial which is involved, $L^{(1/2\pm\alpha)}_p(v)$,
is of order $p$ in $v$ and thus of order $2p$ in $(u-u_0)$, these wave functions possess
a total of $(2p+1)$ zeros (all of integer degree except for the one at $u=u_0$ which behaves nonanalytically
as $|u-u_0|^{1\mp\alpha}$) as expected of states at an odd excitation level, $n=2p+1$.
However this contrasts with the states at even excitation level, $n=2p$, which are even in the variable $(u-u_0)$
and are thus not constrained in their value at the point $u=u_0$.

Away from the value $\alpha=0$ which corresponds to the degenerate situation of the harmonic oscillator which certainly
is not singular at $(u-u_0)$, from (\ref{eq:sol1}) and (\ref{eq:sol2}) one observes the following situation.
For the sake of the discussion, let us consider the case $\alpha>0$, knowing that the case $\alpha<0$ would
correspond to permuting the roles of the two systems $H_1$ and $H_2$ in the analysis hereafter.
For $\alpha>0$, the potential $V_1$ is unbounded below
at $u=u_0$, while $V_2$ is bounded below but diverges to $+\infty$ at $u=u_0$, both potentials behaving as
$(u-u_0)^{-2}$ in the vicinity of that point. Then all wave functions $\psi_{1,2p}(u)$ are seen to vanish
as $|u-u_0|^\alpha$ at $u=u_0$, while all wave functions $\psi_{2,2p}(u)$ diverge at that point as $|u-u_0|^{-\alpha}$.
Yet this divergence is still just mild enough for these latter wave functions as well to be normalisable over $\mathbb{R}$.
In other words, by vanishing at $u=u_0$, the states $\psi_{1,2p}(u)$ avoid falling ``at the bottom" of the infinite negative throat-like
potential well since then their probability density, $|\psi_{1,2p}(u)|^2$, vanishes at the point $u=u_0$. On the other hand for the states
$\psi_{2,2p}(u)$, their probability density diverges as $|u-u_0|^{-2\alpha}$ at $u=u_0$ with a power,
$(-1<-2\alpha<0)$, just mild enough for the total probability to remain finite (and thus normalisable to unity) and for
having a nonvanishing and finite tunneling probability across the potential barrier in spite of a probability
density which grows infinite just at $u=u_0$,
making it thus possible for these ``even" states to have wave functions of even parity in the variable $(u-u_0)$.
Since the generalised Laguerre polynomials $L^{(-1/2\pm\alpha)}_p(v)$ involved in these wave functions are of
order $p$ in $v$, hence of order $2p$ in the variable $(u-u_0)$, the wave functions $\psi_{2,2p}(u)$ indeed
possess $2p$ zeros (of integer degree), as expected of states at an even excitation level, $n=2p$.
However for the states $\psi_{1,2p}(u)$ we encounter the peculiar situation of $(2p+1)$ zeros, all of
integer degree except for the one at $u=u_0$ which behaves nonanalytically as $|u-u_0|^{\alpha}$. And yet, all these
states are at an even excitation level, $n=2p$.

\section{Conclusions}
\label{SectConclusions}

This work addresses and solves the first two simplest cases of a general programme outlined
in the Introduction, within the context of Supersymmetric Quantum Mechanics and its hierarchies of
integrable quantum Hamiltonians intertwined in pairs. Heretofore in the literature in order to restrict
the structures of such hierarchies, the requirement of ``shape invariance" has been used for intertwined
Hamiltonians. In the present work we suggest to go beyond the next level in a hierarchy, and rather relate
Hamiltonians separated by an arbitrary number of levels. A further restriction even is requiring that
the related Hamiltonian systems be in fact identical, differing only by an overall shift in their energy spectra.
The added advantage of this latter requirement is to induce a periodic structure of finite order $N$ in the hierarchy
related to a choice of the first $N$ gaps in the energy spectra to be repeated then in a periodic fashion, and a choice of
potential energies whose energy eigenfunctions would involve specific classes of orthogonal polynomials.
Through the methods proper to Supersymmetric Quantum Mechanics, new results in the form of generalised
Rodrigues formulae and recursion relations for such classes of orthogonal polynomials would also be generated.

Besides the latter interest of a mathematical physics character, there is also a potential real physics interest to
this approach. Indeed, by specifying the first $N$ gaps of some energy spectrum for a single degree of freedom system,
it should be possible to engineer through such methods a quantum Hamiltonian whose energy spectrum includes
these $N$ first energy levels, then repeated periodically, opening the way towards approximation methods
amenable to numerical techniques as well in order to model actual quantum physical systems.
And as mentioned in the Introduction, one last avenue to be explored further along such ideas is that of
inverse scattering and integrable systems.

In the present work the simple case $N=1$, in which two intertwined Hamiltonians are identical up to the overall
shift in energy, is solved as a warm-up exercise. But the detailed and complete solution of the $N=2$ case
is the actual original content of the paper, making explicit in that case most of the ideas outlined above.
It is established that the class of orthogonal polynomials related to this choice is that of the
generalised Laguerre polynomials. And indeed, new generalised Rodrigues formulae have been identified,
as well as specific recursion relations. At the same time new classes of integrable quantum Hamiltonians with
singular potentials, that generalise the harmonic oscillator potential, have been identified and solved
completely solely using algebraic techniques of Supersymmetric Quantum Mechanics. Their quantum states display some
interesting properties related to the singularity of their potential energies.

The next step in this programme would be to manage solving the coupled Riccati recursion relations for $N=3$,
thereby opening the way in fact to the general case with arbitrary $N$. Provided this is feasible in analytic form
while a numerical approach may always be developed, one should
expect new and interesting results relating to the different directions outlined above.

\section*{Acknowledgements}

The present work is supported in part by a ICMPA-UCL Project between the International Chair in Mathematical Physics
and Applications (ICMPA-UNESCO Chair) at the University of Abomey-Calavi (Cotonou, Benin), and the Institute of Research in
Mathematics and Physics (IRMP) at the Catholic University of Louvain (UCL, Louvain-la-Neuve, Belgium), the latter being the
Institution financing entirely this Project whose purpose is to support the training
of mathematicians and physicists from the Democratic Republic of Congo, in particular at the University of Kinshasa (Kinshasa, DRC).

The work of DBI is supported by a PhD Fellowship held alternatively at both Institutions within the framework of this ICMPA-UCL Project.
The work of JG is supported in part by the Institut Interuniversitaire des Sciences Nucl\'eaires (I.I.S.N., Belgium),
and by the Belgian Federal Office for Scientific, Technical and Cultural Affairs through the Interuniversity Attraction Poles (IAP) P6/11.

\end{document}